\documentclass[doublecol,linenumbers]{epl2}
\usepackage{amsmath}
\usepackage{bm}
\usepackage{verbatim} 
\usepackage[]{graphicx}			
\usepackage{color}			
\newcommand{\be}{\begin{equation}}
\newcommand{\ee}{\end{equation}}
\newcommand{\ben}{\begin{eqnarray}}
\newcommand{\een}{\end{eqnarray}}
\newcommand{\bes}{\begin{subequations}}
\newcommand{\ees}{\end{subequations}}
\newcommand{\bb}{\bibitem}

\newcommand{\bfi}{\begin{figure}}
\newcommand{\efi}{\end{figure}}
\newcommand{\bc}{\begin{center}}
\newcommand{\ec}{\end{center}}
\newcommand{\sech}{\mbox{sech}}
\begin{document}
%%%%%%%%%%%%%%%%%%%%%%%%%%%%%%%%%%%%%%%%%%%%%%%%%%%%%%
\title{Compact Structures in Standard Field Theory}
\author{D. Bazeia\inst{1,2}, L. Losano\inst{1,2}, M. A. Marques\inst{1} and R. Menezes\inst{2,3}}
%%%%%%%%%%%%%%%%%%%%%%%%%%%%%%%%%%%%%%%%%%%%%%%%%%%%%%
\institute{
\inst{1}Departamento de F\'\i sica, Universidade Federal da Para\'\i ba, 58051-970 Jo\~ao Pessoa, PB, Brazil\\ 
\inst{2}Departamento de F\'\i sica, Universidade Federal de Campina Grande, 58109-970 Campina Grande, PB, Brazil and \\ 
\inst{3}Departamento de Ci\^encias Exatas, Universidade Federal da Para\'\i ba, 58297-000 Rio Tinto, PB, Brazil.}

\date{\today}


\abstract{We investigate the presence of static solutions in models described by real scalar field in two-dimensional spacetime. After taking advantage of a procedure introduced sometime ago, we solve intricate nonlinear ordinary differential equations and illustrate how to find compact structures in models engendering standard kinematics. In particular, we study linear stability and show that all the static solutions we have found are linearly stable.}

\maketitle

%%%%%%%%%%%%%%%%%%%%%%%%%%
\section{Introduction} 

This work deals with the existence of compact structures in models described by a single real scalar field in two-dimensional spacetime. The main motivation is to shed further light on the subject and contribute for applications in high energy and in condensed matter physics. 
Since compactons may behave as solitons, from this perspective two distinct facts comes to mind: first, that solitons are directly connected with the interplay between dispersion and nonlinearity \cite{b1,b2}; second, that under the presence of nonlinear dispersion, solitons may acquire spatial profiles with compact support \cite{compact,css}.

In models described by scalar fields, compact structures have been recently studied with generalized kinematics; see e.g., Ref.~\cite{defect}. In this case, nonlinear dispersion is present to play the game and give rise to such compact structures. Compactons have also been studied in \cite{compact1} with standard kinematics but with potentials of the $V$-shaped form. In this case, there is no nonlinear dispersion, but nonlinearity has to enter the game with potentials that engender the so-called V-shaped form.

In relativistic models, the investigations \cite{compact1} go against the suggestion that compactons require nonlinear dispersion. This poses an interesting issue, but it requires the somehow artificial feature, that the potential engenders the V-shaped behavior. For this reason, in this work we further investigate the presence of compact structures in models with standard kinematics. Here, however, instead of resorting to $V$-shaped potentials, we take advantage of a former work \cite{bmm} where we have introduced special potentials, giving rise to double kink solutions. We show below that this feature can be used to get to compact solutions, with energy density that vanish outside a compact space. This is the key issue of the work, which suggests a new route to investigate compact solutions in models with standard dynamics, without relying on nonlinear dispersion. The issue here is that the potentials we investigate engender extra minima, with divergent second order derivative. This makes the associated squared mass divergent at those minima, giving room for the presence of compactons. See the recent work \cite{ck} for further information on compactons in models with standard kinematics. 

To work and solve the intricate nonlinear differential equations that appear in such problems, we take advantage of the approach introduced in \cite{com} and we construct new compact structures in models described by a single real scalar field in $(1,1)$ spacetime dimensions. We start from the standard $\chi^4$ model, under the presence of spontaneous symmetry breaking, and we deform it according to the recipe of \cite{com} to get to the new model, which may engenders compact structures.

To make the investigation pedagogical, we organize the work as follows: in the next section we review the procedure introduced in Ref.~\cite{com}, and we introduce and solve two distinct models, finding structures that have kink, half-compact and compact features. We then study stability, showing that the solutions we have found are all linearly stable. We end the work with some comments and conclusions.

%%%%%%%%%%%%%%%%%%%%%%%%%%%%%%%%%%%
\section{The procedure}\label{sec2}

We start the investigation with the Lagrange density for the standard model, with a single real scalar field $\chi=\chi(x,t)$ in $(1,1)$ spacetime dimensions, using dimensionless field, space and time coordinates, and coupling constants, for simplicity. The metric is $(+,-)$ and the Lagrange density
${\cal L}(\chi,\partial_\mu\chi)$ has the form
\be\label{sm}
{\cal L}(\chi,\partial_\mu\chi)=\frac12\partial_\mu\chi\partial^\mu\chi-U(\chi),
\ee
where 
\be\label{chi4}
U(\chi)=\frac12 (1-\chi^2)^2.
\ee
This is the $\chi^4$ model, with spontaneous symmetry breaking. It engenders the kinklike solution $\chi(x)=\tanh(x)$, with energy density $\rho(x)={\rm sech}^4(x)$ and energy $E=4/3$.

%%%%%%%%%%%%%%%%%%%%%%%%%%%%%%%%%%%%%%%%%%
\begin{figure}[t]
\begin{center}
\includegraphics[{width=7cm,angle=-00}]{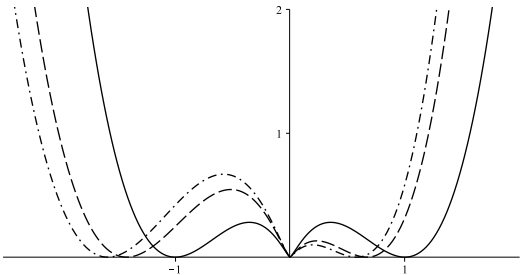}
\end{center}
\caption{The potential of the model 1, for $p=3$ and for $a=0$ (solid line),  $a=1/8$ (dashed line), and $a=1/6$ (dotted-dashed line).}\label{fig1}
\end{figure}
%%%%%%%%%%%%%%%%%%%%%%%%%%%%%%%%%%%%%%%%%%

%%%%%%%%%%%%%%%%%%%%%%%%%%%%%%%%%%%%%%%%%%
\begin{figure}[t]
\begin{center}
\includegraphics[{width=7cm,angle=-00}]{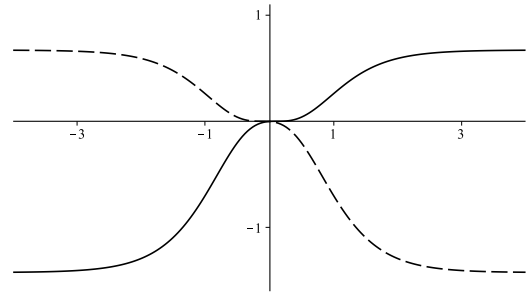}
\end{center}
\caption{The solutions (\ref{dk}), depicted for $p=3$ and $a=1/8$.}\label{fig2}
\end{figure}
%%%%%%%%%%%%%%%%%%%%%%%%%%%%%%%%%%%%%%%%%%

The deformation procedure introduced in \cite{com} shows how to introduce another model, described by the new field $\phi=\phi(x,t)$ and characterized by the new Lagrange density
\be
{\cal L}(\phi,\partial_\mu\phi)=\frac12\partial_\mu\phi\partial^\mu\phi-V(\phi)\label{nm}
\ee
where $V(\phi)$ is the new potential, which specifies the new model. The key ingredient here
is an invertible function $f=f(\phi),$ from which we link the
model (\ref{sm}) with the new model (\ref{nm}). This is done
relating the two potentials $U(\chi)$ and $V(\phi)$ in the very
specific form
\be\label{def}
V(\phi)=\frac{U(\chi\to f(\phi))}{(df/d\phi)^2}.
\ee
Here  $U(\chi\to f(\phi))$ means that in the potential $U(\chi)$ one changes $\chi$ for $f(\phi)$, making it a function of the new field $\phi$. This allows showing that if the starting model has static
solution $\chi(x)$ which obeys the equation of motion
\be
\frac{d^2\chi}{dx^2}=\frac{dU}{d\chi},
\ee
then, the new model has static solution given by 
\be
\phi(x)=f^{-1}(\chi(x)),
\ee
which obeys
\be\label{eom}
\frac{d^2\phi}{dx^2}=\frac{dV}{d\phi}.
\ee
The proof was already given in Ref.~{\cite{com}}.

We recall here that although we are working with static solutions, we can easily boost the solutions to obtain the corresponding traveling waves very naturally, so we omit them here. Also, in $(1,1)$ spacetime dimensions we can write a topological current, in the form
$J_T^\mu=\varepsilon^{\mu\nu}\partial_\nu\phi$,
which is conserved due to the presence of the Levi-Civita (anti-symmetric, pseudo) tensor $\varepsilon^{\mu\nu}$. The topological change
$Q_T\!=\!\phi(x\!\to\!\infty)-\phi(x\!\to\!-\infty)$ is conserved. This is a global property which shows that the nontrivial solutions are protected by the boundary conditions that have to be attained asymptotically. The Derrick scaling argument can also be used, but here it works standardly, leading to global stability of the solutions; see, e.g., Ref.~\cite{bmm}.
  
After this brief review of the procedure, let us now consider some specific functions $f=f(\phi)$ to give rise to new models, together with the respective solutions.

%%%%%%%%%%%%%%%%%%%%%%%%%%%%%%%%%%
\subsection{Model 1}

We first consider the deformation function
\be\label{def1}
f_1(\phi)=a+\phi^{1/p}\,,
\ee
where $0\leq a<1$ and $p=3,5,7,...\,$.
The potential of the new model is given by
\be\label{v1}
V=\frac12p^2\phi^{2-2/p}\left(1-(a+\phi^{1/p})^2\right)^2,
\ee
which has the three minima: $\bar\phi_1=-(1+a)^p,\, \bar\phi_2=0$, and $\bar\phi_3=(1-a)^p$. We depict the potential in Fig.~\ref{fig1}, for some values of the parameters $a$ and $p$. The equation of motion can be obtained easily, from \eqref{eom} and \eqref{v1}; to save space, we do not write it here.

%%%%%%%%%%%%%%%%%%%%%%%%%%%%%%%%%%%%%%%%%%
\begin{figure}[ht]
\begin{center}
\includegraphics[{width=6cm,angle=-00}]{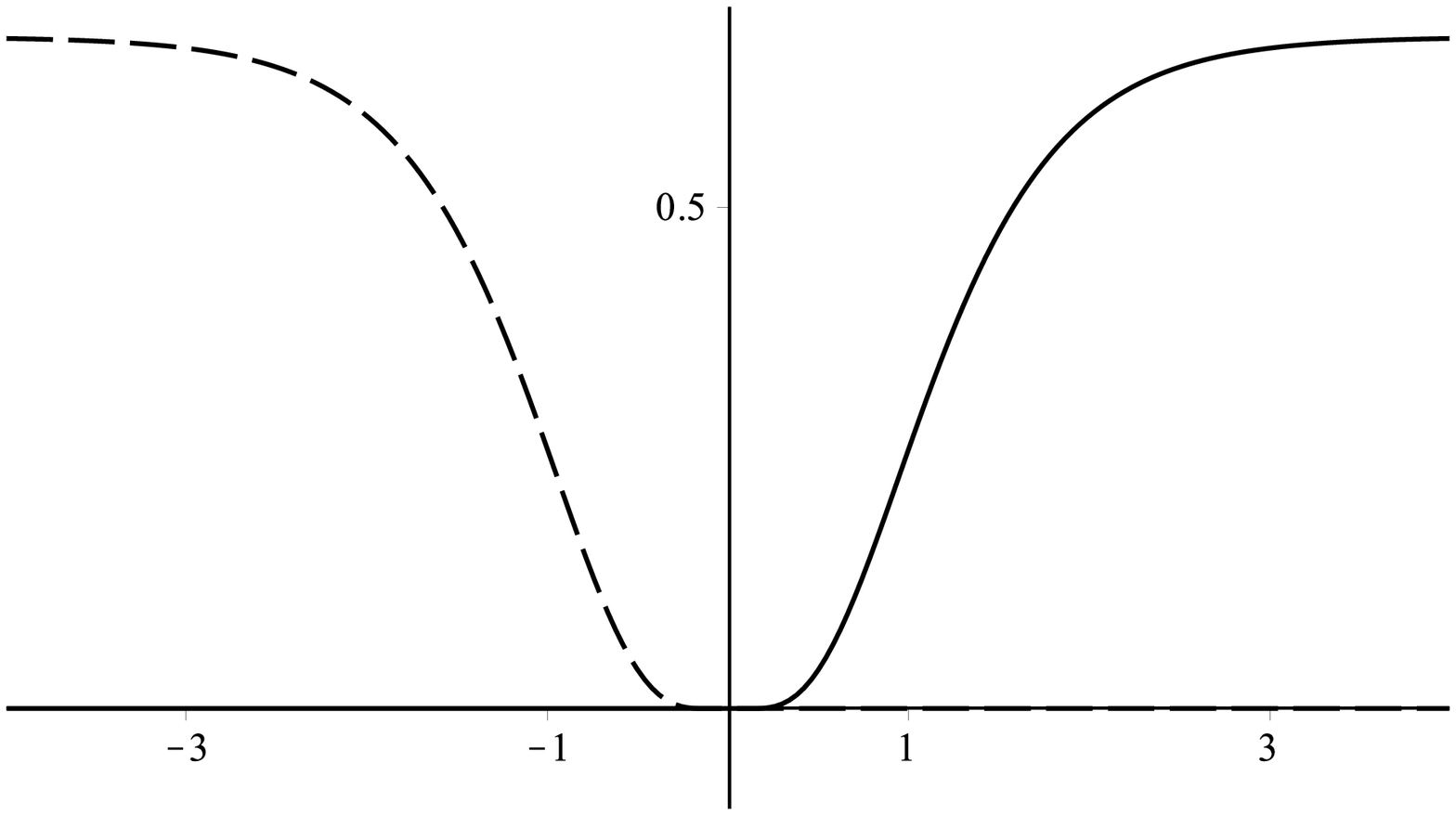}
\includegraphics[{width=6cm,angle=-00}]{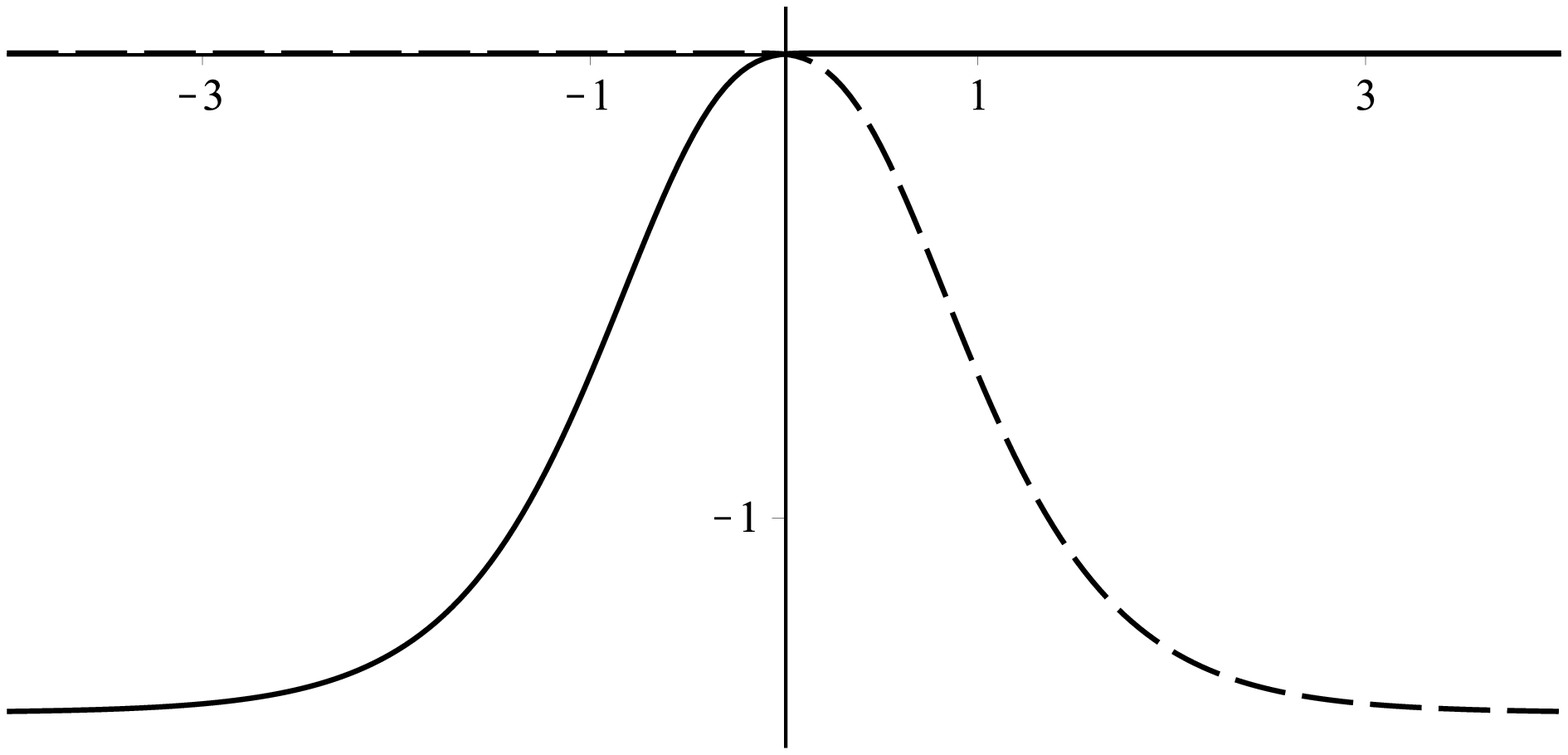}
\end{center}
\caption{The solutions (\ref{hc1}) (bottom) and \eqref{hc2} (top), which are half-compactons, depicted for $p=3$ and $a=1/8$.}\label{fig3}
\end{figure}
%%%%%%%%%%%%%%%%%%%%%%%%%%%%%%%%%%%%%%%%%%

We use this model and the inverse of the deformation function (\ref{def1}) to obtain double kinks, connecting the minima $\phi_1$ and $\phi_3$. They are given by
\be\label{dk}
\phi_{\pm}(x)=-\left(a\pm\tanh(x)\right)^p.
\ee
We depict such structures for specific values of $a$ and $p$ in Fig.~\ref{fig2}. We note that for $a=0$, the model was
introduced in \cite{bmm}, and engenders the $Z_2$ symmetry, with the solutions $\tanh^p(x)$. It is interesting to see that solutions of this type appear in magnetic materials, in the case of specific constrained geometries \cite{mm}. In this sense, the potential (\ref{v1}) describes more general situations for $a\neq0$, and may be of good use to describe more general problems; see, e.g., \cite{DW} and references therein for more recent investigations concerning domain walls in nanoscale ferromagnetic elements.

We note that, for $\bar x={\rm arctanh}(a)$ the above solutions $\phi^{\prime}_{-}(x)=\phi_{-}^{\prime\prime}(x)=0$ at $x=\bar x$, 
and $\phi^{\prime}_{+}(x)=\phi_{+}^{\prime\prime}(x)=0$ at $x=-\bar x$; also, $V^{\prime}=0$ at $\phi=0$. This makes it possible to cut the solutions (\ref{dk}), $\phi_{-}$ in $x=\bar x$ and $\phi_{+}$ in $x=-\bar x$ in order to obtain half-compactons, which also solve the corresponding equation of motion. Explicitly, connecting the minima $\phi_1$ and $\phi_2$ we have 
\bes\label{hc1}
\begin{eqnarray}
\phi^{hc}_{1-}(x) &=& \left\{
\begin{array}{ll}
-\left(a-\tanh(x)\right)^p; \;\;x\leq \bar x,\\ 
0; \;\; x>\bar x,
\end{array} \right.
\end{eqnarray}
and 
\begin{eqnarray}
\phi^{hc}_{1+}(x) &=& \left\{
\begin{array}{ll}
0; \;\;x<-\bar x,\\ 
-\left(a+\tanh(x)\right)^p; \;\; x\geq -\bar x,
\end{array} \right.
\end{eqnarray}
\ees
Also, connecting the minima $\bar\phi_2$ and $\bar\phi_3$ we have
\bes\label{hc2}
\begin{eqnarray}
\phi^{hc}_{2-}(x) &=& \left\{
\begin{array}{ll}
0; \;\;x<\bar x,\\ 
-\left(a-\tanh(x)\right)^p; \;\; x\geq \bar x,
\end{array} \right.
\end{eqnarray}
and 
\begin{eqnarray}
\phi^{hc}_{2+}(x) &=& \left\{
\begin{array}{ll}
-\left(a+\tanh(x)\right)^p; \;\;x\leq -\bar x,\\ 
0; \;\; x>-\bar x.
\end{array} \right.
\end{eqnarray}
\ees

We learn from the above results that the vanishing of the first and second derivative of the solutions and the vanishing of the first derivative of the potential make it possible to cut the solutions and introduce the half-compact features. We get inspiration from this result to propose another model, which we investigate below.

%%%%%%%%%%%%%%%%%%
\subsection{Model 2}

\begin{figure}[ht]
\begin{center}
\includegraphics[{width=7cm,angle=-00}]{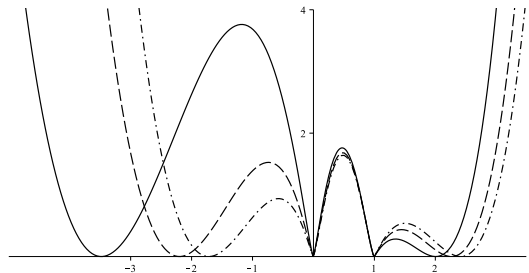}
\end{center}
\caption{The potential of the model 2, for $p=q=3$ and for $b=0.35$ (solid line), $b=0.32$  (dashed line), and $b=0.30$ (dotted-dashed line).}\label{fig4}
\end{figure}

Let us now consider another function, given by
\be\label{def2}
f_2(\phi)=b-(1-\phi^{1/p})^{1/q}\,,
\ee
where $0<b<1$, $p=3,5,7,...$, and $q=3,5,7,...$. In this case the new potential has the form
\be\label{v2}
V=\frac12p^2q^2\phi^{2-2/p}\left(-1+\phi^{1/p}\right)^{2-2/q}\left(1-f_2(\phi)^2\right)^2.
\ee
It contains the four minima: $\bar\phi_1= -((1+b)^q-1)^p, \bar\phi_2= 0,  \bar\phi_3=1$, and $\bar\phi_4=((1-b)^q+1)^p$. We depict this potential for some values of $b$, $p$ and $q$ in Fig.~\ref{fig4}. The equation of motion can be obtained easily, from \eqref{eom}, \eqref{def2} and \eqref{v2}; to save space, we do not write it here.

%%%%%%%%%%%%%%%%%%%%%%%%%%%%%%%%%%%%%%%%%%
%%%%%%%%%%%%%%%%%%%%%%%%%%%%%%%%%%%%%%%%%%
%%%%%%%%%%%%%%%%%%%%%%%%%%%%%%%%%%%%%%%%%%
\begin{figure}[ht]
\begin{center}
\includegraphics[{width=7cm,angle=-00}]{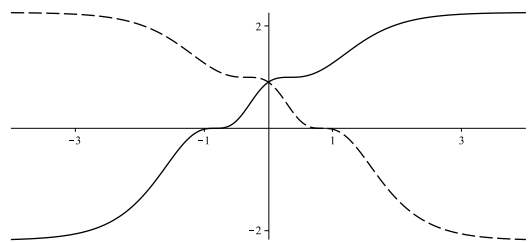}
\end{center}
\caption{The solutions (\ref{tk}), which are triple kinks, depicted for $p=q=3$ and $b=0.32$.}\label{fig5}
\end{figure}
%%%%%%%%%%%%%%%%%%%%%%%%%%%%%%%%%%%%%%%%%%
%%%%%%%%%%%%%%%%%%%%%%%%%%%%%%%%%%%%%%%%%%
\begin{figure}[h]
\begin{center}
\includegraphics[{width=7cm,angle=-00}]{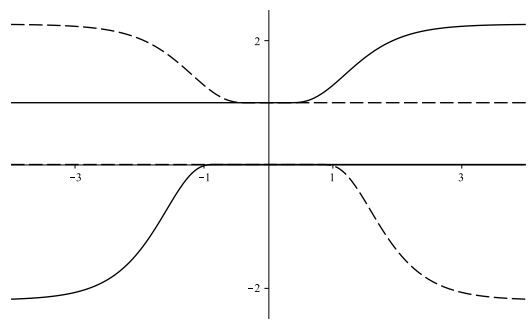}
\end{center}
\caption{The solutions (\ref{hc3}) (bottom) and (\ref{hc4}) (top), which are half-compactons, depicted for $p=q=3$ and $b=0.32$.}\label{fig6}
\end{figure}
%%%%%%%%%%%%%%%%%%%%%%%%%%%%%%%%%%%%%%%%%%
%%%%%%%%%%%%%%%%%%%%%%%%%%%%%%%%%%%%%%%%%%
\begin{figure}[h]
\begin{center}
\includegraphics[{width=7cm,angle=-00}]{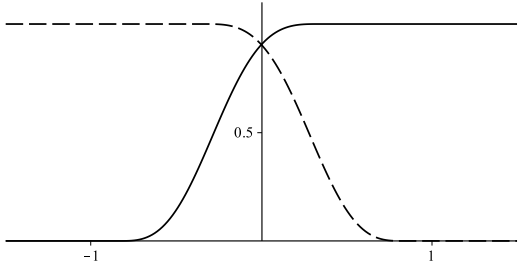}
\end{center}
\caption{The solutions (\ref{ck}), which are compact solutions, depicted for $p=q=3$ and $b=0.32$.}\label{fig7}
\end{figure}
%%%%%%%%%%%%%%%%%%%%%%%%%%%%%%%%%%%%%%%%%%

We take the inverse of the function (\ref{def2}) in order to obtain static solutions connecting the minima $\bar\phi_1$ and $\bar\phi_4$. They have the form of triple kink and antikink, and are explicitly given by
\be\label{tk}
\phi_{\pm}(x)=\left(1-(b\pm\tanh(x))^q\right)^p,
\ee
which we depict in Fig.~\ref{fig5}, for some values of the parameters.

We consider
$\bar x_1={\rm arctanh}(b)$ and $\bar x_2={\rm arctanh}(1-b)$, and we see that $\phi^{\prime}_{-}(x)=\phi_{-}^{\prime\prime}(x)=0$ for $x=\bar x_1$ and $x=-\bar x_2$, and $\phi^{\prime}_{+}(x)=\phi_{+}^{\prime\prime}(x)=0$ for $x=-\bar x_1$ and $x=\bar x_2$. Also, we have that $V^{\prime}=0$ at $\phi=0$ and
at $\phi=1$. These features can be used to cut the solutions (\ref{tk}), $\phi_{-}$ at $x=\bar x_1$ and $x=-\bar x_2$, and  $\phi_{+}$ at $x=-\bar x_1$ and $x=\bar x_2$, thus obtaining half-compacton and compact solutions, which also solve the corresponding equation of motion. So, we have half-compacton defects connecting the minima $\bar\phi_1$ and $\bar\phi_2$ given by
\bes\label{hc3}
\begin{eqnarray}
\phi^{hc}_{1-}(x) &=& \left\{
\begin{array}{ll}
\left(1-(b-\tanh(x))^q\right)^p; \;\;x\leq -\bar x_2,\\ 
0; \;\; x>-\bar x_2,
\end{array} \right.
\end{eqnarray}
and 
\begin{eqnarray}
\phi^{hc}_{1+}(x) &=& \left\{
\begin{array}{ll}
0; \;\;x\leq \bar x_2,\\ 
\left(1-(b+\tanh(x))^q\right)^p; \;\; x\geq \bar x_2.
\end{array} \right.
\end{eqnarray}
\ees
Also, connecting the minima $\bar\phi_3$ and $\bar\phi_4$ they are
\bes\label{hc4}
\begin{eqnarray}
\phi^{{hc}}_{2-}(x) &=& \left\{
\begin{array}{ll}
1; \;\;x<\bar x_1,\\ 
\left(1-(b-\tanh(x))^q\right)^p; \;\; x\geq \bar x_1,
\end{array} \right.
\end{eqnarray}
and 
\begin{eqnarray}
\phi^{hc}_{2+}(x) &=& \left\{
\begin{array}{ll}
\left(1-(b+\tanh(x))^q\right)^p; \;\;x\leq -\bar x_1,\\ 
1; \;\; x>-\bar x_1.
\end{array} \right.
\end{eqnarray}
\ees
In Fig.~\ref{fig6} we depict some solutions for $p=q=3$ and $b=0.32$.

The compact solutions appear connecting the minima $\bar\phi_2$ and $\bar\phi_3$. They are given by
\bes\label{ck}
\begin{eqnarray}
\phi_{-}(x) &=& \left\{
\begin{array}{ll}
0; \;\;x<-\bar x_2,\\ 
\left(1-(b-\tanh(x))^q\right)^p; \;\; -\bar x_2\leq x\leq \bar x_1,\;\;\;\;\;\;\;\;\\ 
1; \;\; x>\bar x_1,
\end{array} \right.
\end{eqnarray}
and
\begin{eqnarray}
\phi_{+}(x) &=& \left\{
\begin{array}{ll}
1; \;\;x<-\bar x_1,\\ 
\left(1-(b+\tanh(x))^q\right)^p; \;\; -\bar x_1\leq x\leq \bar x_2,\;\;\;\;\;\;\;\;\\ 
0; \;\; x>\bar x_2,
\end{array} \right.
\end{eqnarray}
\ees
which we depict in Fig.~\ref{fig7}, for $p=q=3$ and $b=0.32$.

%%%%%%%%%%%%%%%%%%%%%%%%%%%%%%%%%%%%%%%%%%
\begin{figure}[h]
\begin{center}
\includegraphics[{width=7cm,angle=-00}]{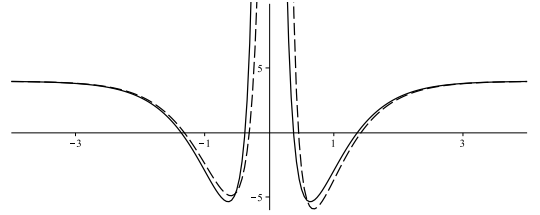}
\end{center}
\caption{The Schroedinger-like potential of the model 1 for the double kink solutions, depicted for $p=3$, $a=0$ (solid line) and  $a=1/8$ (dashed line).}\label{fig8}
\end{figure}
%%%%%%%%%%%%%%%%%%%%%%%%%%%%%%%%%%%%%%%%%%
%%%%%%%%%%%%%%%%%%%%%%%%%%%%%%%%%%%%%%%%%%
\begin{figure}[h]
\begin{center}
\includegraphics[{width=7cm,angle=-00}]{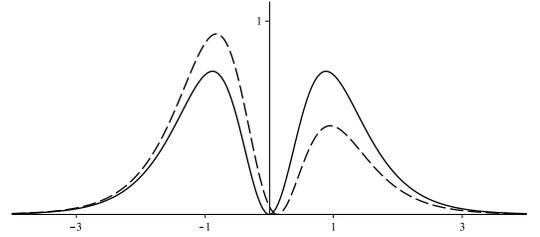}
\end{center}
\caption{The zero mode of the double kink of the model 1, depicted for $p=3$, $a=0$ (solid line) and $a=1/8$ (dashed line).}\label{fig9}
\end{figure}
%%%%%%%%%%%%%%%%%%%%%%%%%%%%%%%%%%%%%%%%%%

%%%%%%%%%%%%%%%%%%%%%%%%%%%%%%%%%%%%%%%%%%
\begin{figure}[t]
\begin{center}
\includegraphics[{width=7cm,angle=-00}]{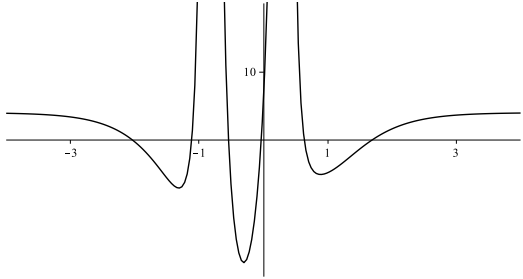}
\end{center}
\caption{The Schroedinger-like potential of the model 2 for the triple kink solutions, depicted for $p=q=3$ and  $b=0.32$.}\label{fig10}
\end{figure}
%%%%%%%%%%%%%%%%%%%%%%%%%%%%%%%%%%%%%%%%%%
%%%%%%%%%%%%%%%%%%%%%%%%%%%%%%%%%%%%%%%%%%
\begin{figure}[t]
\begin{center}
\includegraphics[{width=7cm,angle=-00}]{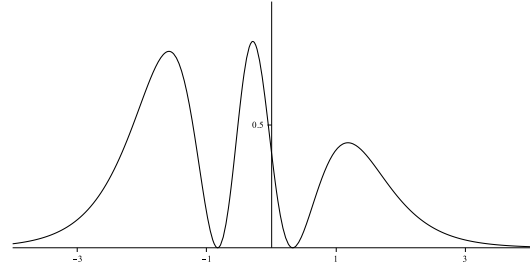}
\end{center}
\caption{The zero mode of the triple kink of the model 2, for $p=q=3$ and  $b=0.32$.}\label{fig11}
\end{figure}
%%%%%%%%%%%%%%%%%%%%%%%%%%%%%%%%%%%%%%%%%%

%%%%%%%%%%%%%%%%%%%%%%%%%%%%%%
\section{Stability}\label{sec3}

Let us now investigate linear stability of the static solutions obtained in the previous
section. Although the solutions are globally stable, protected by the corresponding topological charges, the study of linear stability is in general welcome, and we do it below. In the case of a single field, linear stability requires that one uses
\be
\label{sta}
\phi(x,t)=\phi(x)+\sum_{n}\,\eta_{n}(x)\,\cos(w_{n}t).
\ee
where $\phi(x)$ stands for the static field, and the remaining terms
represent the fluctuations about it. Linear stability implies that the
fluctuations remains limited in time, so that the set of frequencies
$\{w_n\}$ forms a set of real numbers.

We use the field (\ref{sta}) into the second order equation of motion
to obtain the Schroedinger-like equation $H\eta_n(x)=w_n^2\,\eta_n(x)$,
where
\be\label{hamil}
H=-\frac{d^{2}}{dx^{2}}+ U(x),
\ee
and
\be\label{schr}
U(x)=\frac{d^2V}{d\phi^2},
\ee
which must be calculated at the static configuration $\phi(x)$.

An important fact here is that we started from the $\chi^4$ model, with potential that can be written
in terms of $W=W(\chi)$, in the form
\be
U(\chi)=\frac12W^2_\chi,
\ee 
where $W_\chi=dW/d\chi$. In this case, the equation of motion has the form
\be
\frac{d^2\chi}{dx^2}=W_{\chi}W_{\chi\chi},
\ee 
and we can write first-order equation
\be
\frac{d\chi}{dx}=W_{\chi},
\ee
whose solutions solve the equation of motion. For the $\chi^4$ model, from (\ref{chi4}),
we have
\be
W_\chi=1-\chi^2.
\ee
We see that the presence of $W=W(\chi)$ leads to first-order equation, which very much help us to find solutions.

We focus on the new model. We take advantage of (\ref{def}) to write the potential $V(\phi)$ as
\be
V(\phi)=\frac12 \frac{(1-f^2(\phi))^2}{f^{\prime 2}},
\ee
and so we can introduce another function ${\cal W}={\cal W}(\phi)$ such that
\be
{\cal W}_\phi=\frac{1-f^2}{f^{\prime}}.
\ee
It allows that we write the potential $V(\phi)$ as
\be
V(\phi)=\frac12{\cal W}^2_\phi
\ee
Thus, we have
\be
\frac{d^2V}{d\phi^2}={\cal W}^2_{\phi\phi}+{\cal W}_\phi{\cal W}_{\phi\phi\phi}.
\ee
In the new model, the static field obeys
\be
\frac{d^2\phi}{dx^2}=\frac{dV}{d\phi}={\cal W}_{\phi}{\cal W}_{\phi\phi};\;\;\;\frac{d\phi}{dx}={\cal W}_{\phi}
\ee
These results allow that we write the Hamiltonian (\ref{hamil}) in the form
$H=S^{\dag}S$, where
\be
S^\dag=\frac{d}{dx}+ {\cal W}_{\phi\phi}\,,\,\,\,S=-\frac{d}{dx}+{\cal W}_{\phi\phi}.
\ee
This factorization ensures that the Hamiltonians $H$ is positive definite: the corresponding eigenvalues cannot be negative,
thus ensuring linear stability of the static solutions.

This is general result, and now we investigate the two models studied previously, to show how the stability works explicitly.

%%%%%%%%%%%%%%%%%%%%%%%%%%%%
\subsection{Model 1}

For $a=0$, the Schrondiger-like potential (\ref{schr}) has the form
\be\label{schr1}
U(x)=2-20\,\sech^2(x)+2\tanh^{-2}(x).
\ee
In the more general case with $a\neq 0$, the expression is awkward and we do not show it here. Instead, in Fig.~\ref{fig8} we depict the potential for $a=0$ and for $a=1/8$, for the solution $\phi_{-}(x)$ of Eq.~(\ref{dk}). Moreover, the eigenstate for the zero mode of the same double kink of (\ref{dk}) is given by
\be\label{zm0}
\eta_0(x)= p\,(a-\tanh(x))^{p-1} \sech^2(x).
\ee
For the half-compacton (\ref{hc1}), we have
\begin{eqnarray}\label{zm1}
\eta_0^{hc}(x) &=& \left\{
\begin{array}{ll}
p\,(a-\tanh(x))^{p-1} \sech^2(x); \\ {\rm for}\;x\leq \bar x,\\ 
0; \;{\rm for}\; x>\bar x.
\end{array} \right.
\end{eqnarray}
They have no nodes, showing stability of the double kink and half-compacton; see Fig.~\ref{fig9}, where we depict the (normalized) zero mode of the double kink solution.

%%%%%%%%%%%%%%%%%%%%%%%%%%%%%%%%%%
\subsection{Model 2}

In this case, the potential is also awkward, so in Fig.~\ref{fig10} we depict it for $p=q=3$ and $b=0.32$, for $\phi_{-}(x)$ of Eq.~(\ref{tk}). The eigenstate for the zero mode of the same triple kink of (\ref{tk}) is given by
\ben\label{zm3}
\eta_0(x)\!&=& \!pq\,(1\!+\!(-b\!+\!t(x))^q)^{p-1} (-b\!+\!t(x))^{q-1}\!s^2(x),\;\;\;\;\;
\een
where we are using $t(x)=\tanh(x)$ and $s(x)={\rm sech}(x)$, for simplicity. For the half-compacton (\ref{hc3}) we have
\begin{eqnarray}
\eta_0^{hc}(x)\!\! &=&\!\! \left\{
\begin{array}{ll}
pq\,(1+(-b+t(x))^q)^{p-1} (-b+t(x))^{q-1}s^2(x);\nonumber\\
{\rm for} \;x\leq -\bar x_2,\\ 
0; \;\; {\rm for}\;x>-\bar x_2.
\end{array} \right.
\end{eqnarray}
For the half-compacton (\ref{hc4}) one gets
\begin{eqnarray}
\eta_0^{hc}(x)\!\! &=&\!\! \left\{
\begin{array}{ll}
0; \;{\rm for}\;x<\bar x_1\\ 
pq\,(1+(-b+t(x))^q)^{p-1} (-b+t(x))^{q-1}s^2(x);\nonumber\\
{\rm for}\; x\geq \bar x_1,
\end{array} \right.
\end{eqnarray}
and for the compacton \ref{ck} we can write
\begin{eqnarray}
\eta_0^{c}(x)\!\! &=& \!\!\left\{
\begin{array}{ll}
0; \;{\rm for}\;x<-\bar x_2,\\ 
pq\,(1+(-b+t(x))^q)^{p-1} (-b+t(x))^{q-1}s^2(x);\nonumber\\
{\rm for }\; -\bar x_2\leq x\leq \bar x_1,\\ 
0; \;{\rm for}\; x>\bar x_1.
\end{array} \right.
\end{eqnarray}
They have no nodes, showing stability of the triple kink, half-compact and compact solutions; see Fig.~\ref{fig11}, where we depict the (normalized) zero mode of the triple kink solution.

%%%%%%%%%%%%%%%%%%%%
\section{Ending comments}\label{sec4} 

In this work we developed a procedure to construct and solve generalized models described by a single real scalar field in (1,1) spacetime dimensions.
We focused on the presence of static solutions with compact behavior, and we offered distinct models, which support half-compact and compact structures.
The compact behavior appears from the possibility of cutting the solution, making it constant from a given, finite point in the $x$ axis, with no contribution to the energy density. This makes the solution half-compact or compact, depending on the specific features the solution engenders.

To complete the work, we studied linear stability on general grounds, and we showed that the static solutions are all linearly stable. Moreover, we investigated stability of the two models described in this work, identifying the corresponding zero modes and showing that their kink, half-compact and compact solutions are stable.

The procedure presented in this work can be used in a diversity of ways, to help us explore new models and the classical structures they may engender. In particular, the new half-compact and compact structures may suggest new constrained geometries, to make magnetic systems support the new static structures presented in this work. Another issue of current interest concerns the use of compactons in the five-dimensional braneworld scenario, with a single extra dimension of infinite extent. This was recently studied in \cite{ck}, and the several models introduced in the present work may lead us to interesting new scenarios, in particular to the case of an asymmetric brane, induced via half-compactons. This is under investigation, to be reported elsewhere.

After finishing the current work, we became aware of Ref.~\cite{GG}, in which the authors deal with similar issues, investigating compact traveling waves in models with standard kinematics.
  
\acknowledgments We would like to thank G. Gaeta for drawing our attention to Ref.~\cite{GG}. We also thank CAPES and CNPq for partial financial support.

%%%%%%%%%%%%%%%%%%%%%%%%%%%%%%%%%%%%%

\end{document}